# Red Shift in a Laboraory Environment.

Yatsunenko Yu, Budagov Ju.

March 3, 2011

Joint Institute for Nuclear Research, Dubna, Russia

**Abstract**

A hypotheses of energy loss for polarization of $e^-e^+$ vacuum by a photon passing interstellar space is considered. An excitation and relaxation of vacuum can't run with speed of light due to very small but finite fraction of $e^-e^+$ pair mass that creates a retardment in recuperation of deposited energy back to photon. This "forgotten" by many photons energy is finally splashed out in real space as a Relic Radiation. An assumption that such energy loss is proportional to a photon energy conforms to Hubble low of Red Shift and experimental data treated as accelerated expansion of Universe. A possibility of an observation of this type energy loss is considered at high-energy accelerators where energy deposition may reach up hundreds MeV in second.

The astronomical Red Shift, usually explained by an expansion of Universe, could also be caused by a loss of energy of photons as they propagate through space.

Earlier ideas about energy loss based on a scattering of photons by interstellar dust were rejected due to absence of a blurring in images of astronomical objects. Such an absence implies there is no change in the transverse component of photon's momentum in a classical conception of energy loss and therefore no energy loss at all.

There could be another mechanism of the energy loss by a photon passing through an ideal vacuum: let's assume a photon partially polarizes the electromagnetic component of the vacuum, such as it is seen in the Lamb Shift[1]. This excitation of vacuum takes a very small but finite time because of tiny fraction of $e^-e^+$ masses are coming into being and after are being collapsed not immediately – obviously if non-zero mass appears on a scene thereat all the processes can't run with a speed of light. During time of this excitation and relaxation of vacuum photon moves for a distance at speed of light, any energy deposited by this photon for such a polarization of vacuum can't be returned back to the photon, it has already flown away from this scene and none can run down this photon. Thus, such a delay generates a mechanism of energy loss by a photon and apparently only the longitudinal momentum is changed (as it is an axis of "local symmetry") .

This small portion of energy "forgotten" by the photon can be dissipated and accumulated inside e⁻e⁺ vacuum as an excitation and then can finally be emitted in real space as a radio-frequency treated presently as "Relic Radiation"[2][1]

Thus these two constituents of hypothesis do not contravene to energy conservation law, at least.

In a frame of this hypothesis, it is essential to suppose that the photon energy deposition heating up the $e^-e^+$ vacuum is proportional to the energy of photon $h\nu$ with a constant of proportionality A:

$$-d(h\nu) = (h\nu) \cdot A \cdot dt \tag{1}$$

which results in

$$\nu = \nu_o exp(-A \cdot t) \tag{2}$$

The expansion of the universe is described in astrophysics by a Red Shift parameter "z" [2] for the Doppler shift of a photon with frequency $\nu$

$$z = \frac{\Delta \lambda}{\lambda_o} \equiv \frac{\nu_o - \nu}{\nu} \tag{3}$$

The expression of "z" is simple in a Fizeau-Doppler approximation [3] for the relativistic
velocity "v" of an object emitting photons:

$$z \approx \frac{v}{c} \tag{4}$$

("c" is the speed of light, as usual).

The exponential photon energy loss in equation (2) being applied to equation (4) determines
a dependence for z(t) :

$$z = exp(A \cdot t) - 1 \tag{5}$$

First order approximation of exponent gives:

$$z \approx A \cdot t \tag{6}$$

conforming to the Hubble law [2] v = $H \cdot (c \cdot t)$ or:

$$z = H \cdot t \tag{7}$$

Thus one can suggest that "A" (2) is the Hubble constant $H = 2.5 \cdot 10^{-18} s^{-1}$.

---

[1] This Relic Radiation explained as a result of Big Band, has however all the properties of classical radiation of "black body" heated to 2.73°K ... there is one candidate for this "black body" -vacuum

Second approximation of (5) corresponds to ideas of accelerated expansion of Universe:
$$z \approx A \cdot t + \frac{1}{2}(A \cdot t)^2 \qquad (8)$$

Experimental observation of such an acceleration in 1998 [4] for far outlying objects determines z=0.5 this value of z corresponds to almost the middle of Universe : $t_{z=0.5}= 0.4 \cdot T_U$ ($T_U \equiv 1/H$, in assumption A=H ).

Such reasonable numbers bear evidence again for a role of the Hubble constant in hypotheses of energy loss (2), thus one may rewrite equation (2) with the Hubble constant:
$$\nu = \nu_o exp(-H \cdot t) \qquad (9)$$

It would be very interesting to detect this change of photon frequency in experimental conditions for instance with lasers. However a very small scale of energy deposition would be expected at the level of a "Hubble Quanta" (could one give such a name?), H·h =$1.7 \cdot 10^{-39}$ MeV, which can't be apparently seen with a present technology.

One can suggests however that we can see a changing in "Relic Radiation" intensity "around" huge laser in a space (ideal background conditions) detecting by radio telescopes , or more realistic tasks: to try to detect these changing in "Relic Radiation" nearby stellar systems transferring very-high-speed gases (photons, hydrogen) - dark hole is a good candidate.

One can consider that very-high-speed gases of hydrogen or relativistic protons can propagate through the vacuum in a similar way as photon does (nota bene Louis-Victor-Pierre-Raymond,$7^{th}$duke De Broigle's waves !) and the proton energy loss into vacuum can be described by an expression apparently close enough to exponential law (2,9).

Again, each proton emits very little energy, however a bunch of $10^{13}$ protons in the FNAL Tevatron at 1 TeV can produce about 25 MeV/sec of entire energy deposition. For gold atoms accelerated in BNL with $N_{in}$ ~ $10^{12}$ the total energy deposition can be 85MeV/sec and in the LHC ( E=7 TeV, $N_{in}$~$4.7 \cdot 10^{14}$) one could see 800 MeV/sec [5] . This scale of energy looks acceptable for a registration.

How to detect this energy is a matter of future studies; one can point out a few directions: a detection "Relic Radiation" intensity nearby accelerators or a searching for a change in the behavior of heavy radioactive atoms that have a wide spectra of available states for orbital electrons and nuclear decays, again, near accelerators.

One must also notice that whereas Hubble Quanta is very weak, "Relic Radiation" quanta is about $\sim 10^{-3}$eV, thus all the accelerating by LHC, for instance, protons are able to induce $8 \cdot 10^9$in second of 200-300 GHz "Relic Radiation" photons (in coherent production, of course).

There are of course several tasks for detailed development of a proposal for an experiment, however the result of this experiment may open an interesting program for a creation of model of the vacuum.

It is our pleasure to express thanks to Dr. Brendan Casey and Dr. Vladimir Sirotenko for interesting discussions.